\documentclass[epj]{svjour}
\usepackage{graphics}
\usepackage{amsmath}
\usepackage{pdfpages}
\begin{document}
\title{Quantitative predictions on auxin-induced polar distribution of PIN proteins during vein formation in leaves}
\author{Karen Alim \and Erwin Frey}               
\institute{Arnold Sommerfeld Center for Theoretical Physics and Center for NanoScience, Ludwig-Maximilians-Universit\"{a}t, Theresienstr. 37, D-80333 M\"unchen, Germany}
\date{Received: date / Revised version: date}
\abstract{
The dynamic patterning of the plant hormone auxin and its efflux facilitator the PIN protein are the key regulator for the spatial and temporal organization of plant development. In particular auxin induces the polar localization of its own efflux facilitator. Due to this positive feedback auxin flow is directed and patterns of auxin and PIN arise. During the earliest stage of vein initiation in leaves auxin accumulates in a single cell in a rim of epidermal cells from which it flows into the ground meristem tissue of the leaf blade. There the localized auxin supply yields the successive polarization of PIN distribution along a strand of cells. We model the auxin and PIN dynamics within cells with a minimal canalization model. Solving the model analytically we uncover an excitable polarization front that triggers a polar distribution of PIN proteins in cells. As polarization fronts may extend to opposing directions from their initiation site we suggest a possible resolution to the puzzling occurrence of bipolar cells, such we offer an explanation for the development of closed, looped veins. Employing non-linear analysis we identify the role of the contributing microscopic processes during polarization. Furthermore, we deduce quantitative predictions on polarization fronts establishing a route to determine the up to now largely unknown kinetic rates of auxin and PIN dynamics.
\PACS{
      {87.17.Pq}{Morphogenesis}   \and
      {87.10.Ca}{Analytical theories}    \and
      {87.18.Hf}{Spatiotemporal pattern formation in cellular populations}  \and
      {82.40.Bj}{Oscillations, chaos, and bifurcations}
     } 
}
\authorrunning{K.~Alim and E.~Frey}
\titlerunning{Auxin-induced polar distribution of PIN proteins}
\maketitle
%
%%%%%%%%%%%%%%%%%%%%%%%%%%%%%%%%%%%%%%%%%%%%%%%%
\section{Introduction}
%%%%%%%%%%%%%%%%%%%%%%%%%%%%%%%%%%%%%%%%%%%%%%%%
The polar transport of the plant hormone auxin is the key regulator of many processes in the spatial and temporal organization of development and growth of plants. As the indole-3-acetic acid, in short auxin, induces the polar localization of its own efflux facilitator, a member of the family of PIN proteins, a variety of auxin and PIN patterns arise \cite{Petrasek:2009p4589}. Those distributions change dynamically as plants orient in response to environmental stimuli denoted tropism \cite{Friml:2002p1835,Marchant:1999p4592}. During the morphogenesis of plants PIN and auxin rearrangements lie at the heart of organ positioning via phyllotaxis \cite{Reinhardt:2003p348} and vein patterning in leaves \cite{Sieburth:1999p3331}.  

The notion that auxin is transported in a polar, directed, manner inspired researchers since its discovery by Went in 1933 \cite{Went:1933p3768}.  Early works already suggested the participation of a polar localized efflux carrier in the transport of auxin \cite{RUBERY:1973p3312,Goldsmith:1981p503,MITCHISON:1980p3359}, well before its discovery in the form of membrane bound PIN proteins a decade ago \cite{Galweiler:1998p1908}. Since then numerous experiments confirmed that PIN proteins facilitate the efflux of auxin from cells in plants \cite{Okada:1991p3335,Friml:2002p1835,Benkova:2003p3360,Friml:2003p3325,Reinhardt:2003p348}, yeast and mammalian cells, which had been supplied with auxin and PIN \cite{Petrasek:2006p531}. A feedback between auxin and its efflux facilitator localization was proposed by Sachs in his canalization hypothesis \cite{SACHS:1969p1017}, later formalized by Mitchison \cite{MITCHISON:1980p492,MITCHISON:1981p365}. Canalization predicts a feedback of auxin flow between neighboring cells on the amount of efflux facilitators favoring the direction of auxin flow. Experiments confirmed a definite feedback between auxin and PIN distribution \cite{Sauer:2006p1907,Scarpella:2006p352}, the cause of which is reported to lie in auxin affecting the clathrin dependent endocytotic cycling of PIN \cite{Paciorek:2005p1843,Dhonukshe:2008p367}.  Late investigations also identified biochemical processes taking part in the PIN localization in response to auxin, see Ref.~\cite{Benjamins:2008p381} for a review. 

\begin{figure*}[t]
\centering
\resizebox{0.75\textwidth}{!}{\includegraphics{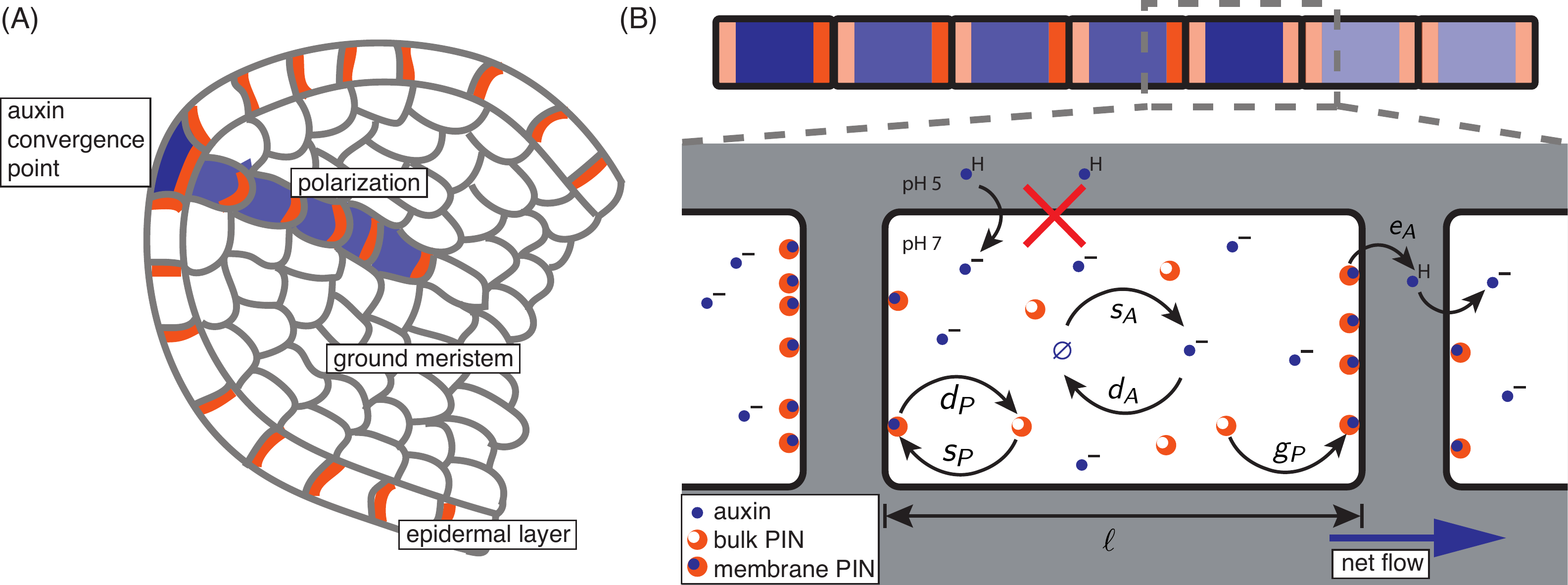}}
\caption{ Illustration of the dynamics of auxin and its efflux facilitator PIN. (A) Schematic drawing of vein initiation in a leaf primordium. Auxin ({\it blue}) accumulates at convergence points in the outer epidermal layer ({\it rim}), from which it is transported into the ground meristem. Due to this inflow of auxin ground meristem cells become polarized in their PIN  distribution ({\it orange}).  (B) Polarization of PIN distribution in a strand of cell due to auxin inflow from the left, indicating details of auxin and PIN dynamics. The weak acid auxin accumulates in the interior of a plant cell due to a gradient in pH. In the inner cell the charged anion is trapped and can only be transported outwards by help of efflux facilitators in the form of PIN proteins. The auxin transport from cell to cell has an efficiency $e_A$. Auxin synthesis $s_A$ and degradation $d_A$ takes place in the inner cell. PIN proteins cycle between the bulk and the cell membrane by basal attachment $s_P$ and detachment rates $d_P$. In addition, positive auxin net flow is modeled to feed back on these rates increasing the PIN attachment by $g_P$. Along the strand of cells the color shading indicates relative concentration of auxin and PIN on either membrane. }
\label{fig_cartoon}
\end{figure*}

A variety of microscopic models for the dynamics of auxin and PIN proteins have been developed to describe their patterns during phyllotaxis \cite{Jonsson:2006p421} and leaf vein formation \cite{Feugier:2005p548}, see  \cite{Berleth:2007p473} for a review. Extensive simulations of these microscopic models describe qualitative aspects of plant development. However, the role of the underlying biological processes and their kinetic rates still remain elusive to a large extent. Quantitative predictions based on analytical solutions of the microscopic equations in a simple scenario might on the one hand help to estimate kinetic parameters and on the other hand give insight into the impact of certain processes.  A scenario amenable to such an investigation is the polarization of in this particular case PIN1 distribution due to auxin flow in the earliest stage of vein formation \cite{Smith:2009p5137}, see fig.~\ref{fig_cartoon}. 
Vein initiation itself takes place in the ground meristem tissue of leaf primordia. The positions of vein initiation sites are determined by auxin accumulation in ``convergence'' points, which lie in a rim of  epidermal cells around the ground meristem tissue \cite{Scarpella:2006p352,Wenzel:2007p1028,Bayer:2009p4590,Koenig:2009p4588}. These single cells with high auxin concentration polarize towards the ground meristem and locally transport their auxin into a cell in the ground meristem tissue. This localized inflow triggers the successive polarization of PIN distributions along a strand of cells starting from the cell with auxin inflow \cite{Scarpella:2006p352}. The strand of polarized cells finally extends up to a previously existing strand of polarized cells, building the pre-pattern for the vascular network. Starting from the petiole of the leaf primordium the polarized cells differentiate then into vascular cells \cite{Sawchuk:2007p4455}. In particular second order veins in {\it Arabidopsis thaliana} exhibit PIN polarization in opposite directions starting from a single bipolar cell, which lies in the ground meristem below the auxin convergence point in the epidermal layer \cite{Scarpella:2006p352}. This yet unresolved behavior gives rise to the formation of closed vein loops when both oppositely polarized strands connect to already formed veins. 

A resolution on the origin of bipolar cells is postulated by examination of a minimal canalization model for the polarization of PIN distribution due to auxin supply in a one-dimensional strand of cells. Performing a non-linear analysis of the model reveals for each single cell two uniform stable states considering polarization. One resting state, where efflux facilitators are symmetrically distributed within the cell, and one polar state characterized by a constant net transport of auxin due to a polar localization of PIN proteins. The model predicts auxin triggered polarization pulses and fronts as a consequence of a dynamic rearrangement of PIN efflux facilitators towards the polar state. Cells with continuous auxin supply can be in a dynamic bipolar state, from which polarization fronts travel to both ends of a strand of cells. The role of the underlying kinetic processes becomes explicit in the course of the non-linear analysis of the polarization. An analytic solution result in quantitative predictions on the pulse's and front's auxin amplitude depending on the kinetic parameters, establishing a basis for detailed experimental determination. 
%%%%%%%%%%%%%%%%%%%%%%%%%%%%%%%%%%%%%%%%%%%%%%%%
\section{Model}
%%%%%%%%%%%%%%%%%%%%%%%%%%%%%%%%%%%%%%%%%%%%%%%%
To describe how auxin polarizes the distribution of PIN during vein formation we focus on a one-dimensional strand of cells, see fig.~\ref{fig_cartoon}~(B), assuming that there is no net auxin flow perpendicular to the direction of polarization. The strand is subdivided into cells of length $\ell$ numbered by $n$. Every cell is characterized by a single auxin concentration $A(n)$ and the concentration of membrane bound PIN proteins. We distinguish between PIN proteins in the bulk of the cell $P_b(n)$ and  adsorbed to the cell membrane either on the right hand side or on the left hand side of the cell, $P_r(n)$, and $P_l(n)$, respectively. The auxin concentration per cell changes due to synthesis and degradation with rates $s_A$ and $d_A$. Furthermore, the amount of auxin changes due to a net flow to neighboring cells facilitated by PIN proteins embedded in the cell membranes of the corresponding cell-cell interface. The net flow from cell $n$ to cell $n+1$ is, hence, given by $J(n)=e_A[A(n)P_r(n)-A(n+1)P_l(n+1)]$, where $e_A$ denotes the transport efficiency rate across the cell-cell interface. The full auxin dynamics is then described by,
\begin{equation}
\frac{d}{dt}A(n)=s_A-d_AA(n)-\frac{1}{\ell}\left[J(n)-J(n-1)\right].
\label{eqn_a}
\end{equation}
 Stating these dynamics for auxin we assume that auxin transport is dominantly from one inner cell to the other. Auxin is known to accumulate in cell interiors \cite{RUBERY:1973p3312,RAVEN:1975p3293} since auxin is a lipophilic weak acid which easily enters cells as undissociated acid, its prevailing form at the pH of 5 in extracellular space. In the interior of plant cells ion pumps keep the pH at 7 leading to the ionization of auxin and building a concentration gradient that accumulates auxin inside cells. As lipid membranes are impenetrable for the now charged molecule, auxin is trapped in the cell's interior rendering efflux facilitator necessary. Being exported by PIN proteins auxin is free to diffuse in the extracellular space, the apoplast. However, the distance  between neighboring cells is so small that almost all auxin molecules have entered any cell again within one millisecond \footnote{The importance of extracellular diffusion of auxin can be assessed by estimating the residence time of auxin in extracellular space. Assuming an auxin molecule diffuses with diffusion constant $D=67\mu m^2/s$  \cite{Swarup:2005p429} in a typical cell-cell interface of $0.5\mu m$. If it comes close to either of the cells the molecule may reenter.   Taking into account that not all extracellular auxin molecules are protonated and hence able to penetrate the membrane we assume the probability to enter a cell to be of $10\%$ \cite{RAVEN:1975p3293,Goldsmith:1981p503}. Considering these assumptions already $97 \%$ of all auxin molecules have reentered any cell including the one they were delivered from after a time duration of one millisecond \cite{USENKO:1987p3447}. }.  Furthermore, auxin can be assumed to be approximately uniformly distributed within typical cells because the diffusion of the small molecule auxin is very large. In aqueous solutions  $D=670\mu m^2/s$ has been measured \cite{byKPaech:1955p3232} which has been confirmed by indirect measurements of the diffusion constant in auxin transport experiments \cite{WANGERMANN:1981p3137}. Distinct auxin gradient would therefore only arise for large plant cells of about $100\mu m$ or larger. It is not entirely clear that this reasoning holds under the condition of highly effective auxin transport \cite{Kramer:2007p401}, however, the time scale of auxin transport by mere diffusion through a cell of typical $50\mu m$ is $4s$, faster than other process contributing to the polarization of PIN, substantiating the neglect of auxin gradients within cells during PIN polarization. Hence, we can approximate auxin flow to be dominated by cell-to-cell transport. Synthesis and degradation of auxin are considered since they take place on fast time scales as the majority of auxin is stored in its conjugated form inside the cell which is readily hydrolyzed in less than seconds \cite{Pollmann:2002p3769}. This is in contrast to the production and degradation of PIN proteins, which takes place on much larger time scales of several minutes. We therefore model the total amount of PIN proteins $P_{\text{tot}}$ to be constant per cell, yielding the following equality for the number of free PIN proteins in the bulk $P_b(n)=P_{\text{tot}}-P_r(n)-P_l(n)$. Hence, we only consider the dynamics of PIN proteins embedded in a cell membrane. Their concentration changes first of all by a basal adsorption rate $s_P$ and  a basal desorption rate $d_P$. Additionally, the net auxin flow over a cell-cell interface is modeled to feed back onto the amount of PIN proteins favoring the flow direction. This is cast in an enhanced attachment or equally a decreased desorption rate: $g_P J^2(n) \theta(J(n))$ as proposed by canalization models \cite{SACHS:1969p1017,MITCHISON:1980p492,MITCHISON:1981p365,Feugier:2005p548,RollandLagan:2005p476,Stoma:2008p4203,Kramer:2009p4196,Smith:2009p5137}. By imposing the Heaviside step function $\theta$, the feedback reacts to positive net flow only. For the following analysis the feedback is proportional to the square of the net auxin flow as stated above, in the discussion we explain that any exponent larger than one yields analogous results. Different feedback mechanisms proposed recently  \cite{Jonsson:2006p421,Smith:2006p585,Merks:2007p3119,Ibanes:2009p4064} are also compared to our approach in the discussion.  Incorporating the positive feedback on auxin flow, the PIN dynamics are given by,
\begin{eqnarray}
\label{eqn_pr}
\frac{d}{dt}P_r(n)&=&-d_PP_r(n)+s_P P_b(n)\nonumber\\
&& +g_PJ^2(n)\theta(J(n))P_b(n),\\
\frac{d}{dt}P_l(n)&=&-d_PP_l(n)+s_PP_b(n)\nonumber\\
&&+g_PJ^2(n-1)\theta(-J(n-1))P_b(n).
\label{eqn_pl}
\end{eqnarray}
Except for the non-linear feedback term, we assumed throughout the model setup linear relationships as a first order expansion to the yet elusive detailed underlying dynamics. We refer to the discussion for an analysis of model modifications confirming the robustness of our assumptions.  In contrast to many existing canalization models  \cite{MITCHISON:1980p492,MITCHISON:1981p365,RollandLagan:2005p476,Stoma:2008p4203,Kramer:2009p4196,Smith:2009p5137} we account for the detailed PIN cycling by endosomes similar to Ref.~\cite{Feugier:2005p548}, however, we discard the explicit dynamics of a putative auxin synthesizer used in that work.

Up to now our  model involves six kinetic rates, however, rescaling the concentration of auxin  $a=A/A_{\text{eq}}$, where  $A_{\text{eq}}=s_A/d_A$, and PIN proteins, $p_{r,l}=P_{r,l}/P_{\text{tot}}$, as well as time $\tau=t\, d_P$ reveals that only four independent, dimensionless parameters govern the behavior of auxin and PIN dynamics, namely
$\delta_a=d_A/d_P$, $\sigma_p=s_P/d_P$, $\gamma_p=g_P A_{\text{eq}}^2P_{\text{tot}}^2/d_P$,  and $\epsilon_a=e_A P_{\text{tot}}/\ell d_P$. Quantitative knowledge of the kinetic rates is very sparse. Half-life measurements of auxin yield estimates for its degradation rate, $d_A=2\cdot10^{-4}-2\cdot10^{-5}1/s$  \cite{Rapparini:2002p523}, which is however strongly affected by environmental conditions such as light, wind, and temperature. Permeability measurements \cite{Delbarre:1996p1372,Swarup:2005p429} of PIN assisted auxin anion transport are found to be $e_A P_{\text{tot}}=1.4\mu m/s$. For the other kinetic rates no experimental estimates are available to best of our knowledge although various rates have been assumed in simulations. This limited knowledge of the kinetic rates underlying auxin and PIN dynamics demonstrates how desired an intuition of their relation and role is, which can be obtained from mathematical analysis, opening up new approaches for experimental measurements. 
%%%%%%%%%%%%%%%%%%%%%%%%%%%%%%%%%%%%%%%%%%%%%%%%
\section{Results}
%%%%%%%%%%%%%%%%%%%%%%%%%%%%%%%%%%%%%%%%%%%%%%%%
\subsection{Observations from numerics}
\begin{figure}[ht]
\centering
\resizebox{0.4\textwidth}{!}{\includegraphics{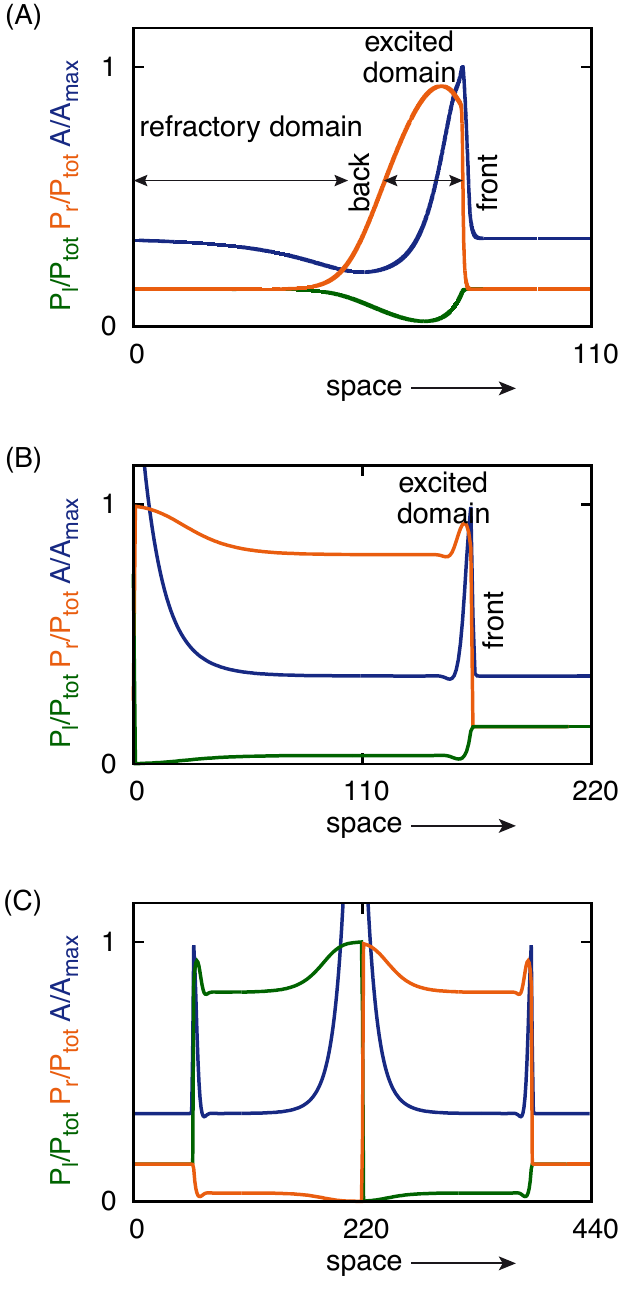}}
\caption{Spatial trajectory of polarization pulse and fronts. Auxin ({\it blue}) and PIN concentration on the left ({\it green}) and right ({\it orange}) hand side of a cell are displayed along a strand of cells numbered by their distance in cells from the site of initiation. (A) A short, initial supply of auxin to a single cell yields a single polarization pulse, while (B) a continuous, high inflow of auxin from one end ({\it left}) yields a polarization front. (C) Continuous supply at a center cell results in two opposite polarization fronts originating from a single bipolar cell. Trajectories arise from the numerical integration of our mathematical model for parameters values $d_A/d_P=0.2$, $s_P/d_P=0.2$, $e_A P_{\text{tot}}/\ell d_P=2$, and $g_PA_{\text{eq}}^2P_{\text{tot}}^2/d_P=8$. Initial and continuous auxin supply of $A=20A_{\text{eq}}$.}
\label{fig_polar}
\end{figure}
During vein formation auxin supplied from the outer epidermis enters a single cell initiating the polarization of a cell strand \cite{Scarpella:2006p352}. We simulated this scenario by integrating the microscopic equations (\ref{eqn_a}-\ref{eqn_pl}) numerically for different kinds of auxin supply, see fig.~\ref{fig_polar}. Starting from a strand of cell with evenly distributed PIN proteins and an equilibrated amount of auxin, we shortly applied auxin by increasing the initial auxin concentration in a single cell.  This triggers a polarization pulse which shortly polarizes the PIN proteins in each cell before all cells relax back to their initial non-polar state, fig.~\ref{fig_polar} (A). Subsequent pulses can only be excited when the cells are almost relaxed back to their non-polar state therefore a certain lag time is required (data not shown). If the auxin is supplied continuously by keeping the amount of auxin in a single cell at high level a polarization front forms. The front causes all cells which it passed to become permanently polarized, fig.~\ref{fig_polar} (B). If a cell in the center of a strand of cells is continuously supplied with auxin two fronts arise traveling to opposite directions along the strand, fig.~\ref{fig_polar} (C). The latter two observations resemble those from vein formation \cite{Scarpella:2006p352}.

As the polarization pulse and front bear a lot of characteristics in common, deriving analytic solutions for the first gives also quantitative insight into the second. In the following our non-linear analysis explains the formation of a polarization pulse and front. Identifying the role of the underlying kinetic processes by exemplarily solving the polarization pulse we derive quantitative results for pulse and front properties. Observations in fig.~\ref{fig_polar}~(A) and (B) indicate that changes in concentration of PIN proteins on the left, $P_l$, facing adverse to the direction of transport, are very small. We therefore assumed in our following analysis $d P_l(n)/dt=0$, i.e., considering the stationary state value $P_l(n)=\sigma_p(1-P_r(n))/(1+\sigma_p)$  for a polarization traveling to the right.

%%%%%%%%%%%%%%%%%%%%%%%%%%%%%%%%%%%%%%%%%%%%%%%%
\subsection{Static state of a single cell}
\begin{figure*}[t]
\centering
\resizebox{0.9\textwidth}{!}{\includegraphics{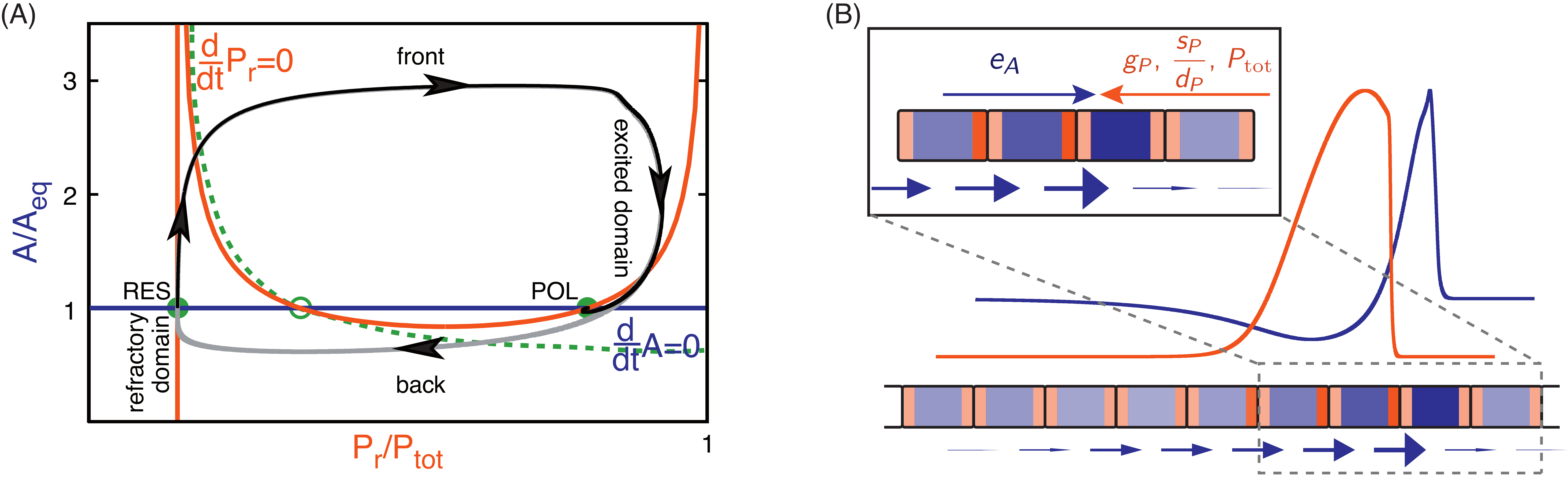}}
\caption{Illustration of the non-linear characteristics of auxin induced polarization. (A) Trajectory of a polarization pulse ({\it grey}) and front ({\it black}) in the nullcline graph obtained from integrating the microscopic equations. Each single cell has two linearly stable fixed points ({\it solid green circle}), a resting state with symmetric PIN distribution ({\it RES}) and a polar state with constant auxin flux ({\it POL}). Their areas of attraction are separated by an unstable manifold ({\it green dashed line}) embedding the unstable fixed point ({\it open green circle}). Parameters as in fig.~\ref{fig_polar}. (B) Heuristic mapping of the numerical polarization pulse along a strand of cells. Auxin is piled up in a cell with even distribution of PIN on both cell membranes at the front of the pulse.  Due to the slow attachment rates of PIN proteins $g_P$, $s_P/d_P$ the cell is still non-polar. Increasing the attachment rates due to the positive feedback of auxin flow reduces the auxin amplitude. Growing cell-to-cell transport efficiencies $e_A$ increase the auxin amplitude as more auxin reaches the peak per time step.}
\label{fig_null}
\end{figure*}
Assuming a uniform state for a whole strand of cells each single cell itself has two stable and one unstable equilibrium state, as shown in fig.~\ref{fig_null}~(A). The first stable fixed point at 
\begin{eqnarray}
a^{\text{RES}}&=&1,\\
p^{\text{RES}}&=&\frac{\sigma_p}{1+2\sigma_p},\nonumber
\end{eqnarray}
 is a resting state, where PIN proteins are evenly distributed and no net auxin flow occurs. For parameters beyond $\gamma_p\geq 4(1+2 \sigma_p)$ two further crossings of the nullclines $d A/dt=0$ and $d P_r/dt=0$ occur in a saddle-node bifurcation, a pair of one unstable and one stable fixed point at 
\begin{eqnarray}
\label{eqn_polar}
a^{\text{POL}\mp}&=&1,\\
 p^{\text{POL}\mp}(a^{\text{POL}\mp})&=&\frac{1+3 \sigma_p\mp(1+\sigma_p)\sqrt{1-\frac{4(1+2 \sigma_p)}{\gamma_p(a^{\text{POL}\mp})^2}}}{2(1+2\sigma_p)},\nonumber
\end{eqnarray}
respectively. At the second stable fixed point the PIN distribution is polar as $P_r$ outnumbers $P_l$ by at least $P_{\text{tot}}/2$, yielding a constant net flow of auxin to the right. The resting state originates from synthesis and degradation terms in eqs.~(\ref{eqn_a}-\ref{eqn_pl}), while the second pair of fixed points arises due to the feedback. As both the resting and the polar state are linearly stable a uniform set of cells decays into one of them depending on the cells initial state. Left of the unstable manifold embedding the unstable fixed point, depicted as dashed line in fig.~\ref{fig_null}~(A), all cells relax to the resting state while right to this separatrix all states decay to the polar fixed point. This is true for a homogeneous set of cells, however, in a spatially inhomogeneous system complex scenarios such as waves and fronts arise \cite{SergeevichZykov:1987p1055,SMikhailov:1991p1044}.   

%%%%%%%%%%%%%%%%%%%%%%%%%%%%%%%%%%%%%%%%%%%%%%%%
\subsection{Dynamic transition}
Due to spatial inhomogeneities passed on along a strand of cells from a cell with auxin supply, the state of a cell changes over time as a polarization pulse or front travels through. The trajectory of states of a single cell in time, fig.~\ref{fig_null}~(A), maps onto the trajectory of a polarization pulse over a strand of cells in space, see fig.~\ref{fig_null}~(B), allowing for a heuristic interpretation of the polarization dynamics. To investigate these dynamics we performed the continuum limit of the microscopic equations (\ref{eqn_a}), (\ref{eqn_pr}), see Supplemental I for details. This gives rise to a set of partial differential equations, which describe the change of $a$ and $p_r$ at one point in space over time, as shown in eqs.~(\ref{eqn_acont},\ref{eqn_prcont}) above.
\begin{figure*}[!t]
\begin{eqnarray}
\label{eqn_acont}
\frac{\partial}{\partial \tau}a(x,t)&=&\delta_a\left(1-a(x,t)\right)-\epsilon_a \frac{1+2 \sigma_p}{1+\sigma_p}\frac{\ell\partial}{\partial x}\left[\left(p_r(x,t)-p^{\text{RES}}\right)a(x,t)\right],\\
\frac{\partial}{\partial \tau}p_r(x,t)&=&-\gamma_p\frac{(1+2 \sigma_p)^2}{(1+\sigma_p)^3}a^2(x,t)\left(p_r(x,t)-p^{\text{RES}}\right)\left(p_r(x,t)-p^{\text{POL}-}(a)\right)\left(p_r(x,t)-p^{\text{POL}+}(a)\right)\nonumber\\
&&-\gamma_p\frac{\sigma_p^2}{(1+\sigma_p)^3}\left(p_r(x,t)-1\right)\left\{\frac{\ell\partial}{\partial x}\left[\left(p_r(x,t)-1\right)a(x,t)\right]\right\}^2.
\label{eqn_prcont}
\end{eqnarray}
\end{figure*} 
This time evolution depends on reaction terms, including only auxin and PIN concentration at the same specific point in space, and gradient terms which account for the influence of neighboring sites.  The reaction terms cause the system to relax to its stable fixed points as described in the previous section. The gradient terms, however, drive the system along its pulse trajectory.  

The following explains the dynamic transition and the role of reaction and gradient terms in it starting from a cell in the non-polar resting state (RES). If the gradient in auxin and PIN to the neighboring cell is large enough the cell is forced out of its stable resting state to larger values of auxin entering the domain of attraction of the polar state (POL). If a neighboring cell has accumulated more auxin and has a higher amount of PIN facing the direction of the polarization pulse, $p_r$, it is very effective in transporting auxin onwards into a cell, raising the auxin content well above the equilibrium value. As the now auxin supplied cell has itself more auxin to transport onwards, the net flux increases starting off the positive feedback which results in PIN polarization. The then fully polarized cell is very efficient in moving its excess auxin onwards, finally decreasing its auxin content towards the polar stable state. The neighboring cell that has been polar and transporting on auxin for a bit longer has less auxin, reversing the direction of the auxin gradient. If this auxin gradient is large enough it drives the cell past its polar state into the domain of attraction of the resting state. Hence, if  the amount of auxin in the neighboring cell is very low, the auxin supply breaks down and with it the onward flux of auxin. The positive feedback is decreasing and with it polarization towards the non-polar resting state. Subsequent polarization pulses can only be triggered if the polarization already saturated down to almost resting state values, otherwise the remaining polarization would just transport the applied auxin onwards before the positive feedback can build up an enhanced polarization of PIN proteins. The phenomenon that a system has to relax back to its resting state before a new pulse can be excited is denoted refractory phase in excitable media.
    
We find that the gradient terms in the auxin and PIN dynamics have unequal analytic structures that lead to their different functions. In the PIN dynamics the squared gradient increases spatial inhomogeneities in PIN distribution by augmenting $p_r$ up to saturation.  In the auxin dynamics the signs of auxin and PIN gradients decide the direction of change in auxin concentration. In front of an excitation pulse the gradient induces a growth of auxin content, while it decreases the amount of auxin at the rear of the pulse. The magnitude of the auxin gradient also decides between the formation of a polarization pulse or front. If the auxin content in the neighboring cell is only slightly smaller due to continuous or only slowly varying auxin supply a constant flux of auxin through polar cells is established, a polarization front forms. If, however, the auxin gradient is large and the auxin inflow decreases drastically the feedback breaks down and the cells in the polarization pulse relax back to their resting state.  In summary the  polarization of PIN distribution by auxin flow is initially a bistable system that behaves like an excitable medium depending on the amount and continuity of auxin supply.

%%%%%%%%%%%%%%%%%%%%%%%%%%%%%%%%%%%%%%%%%%%%%%%%
\subsection{Analytical results}
\begin{figure}[htbp]
\centering
\resizebox{0.4\textwidth}{!}{\includegraphics{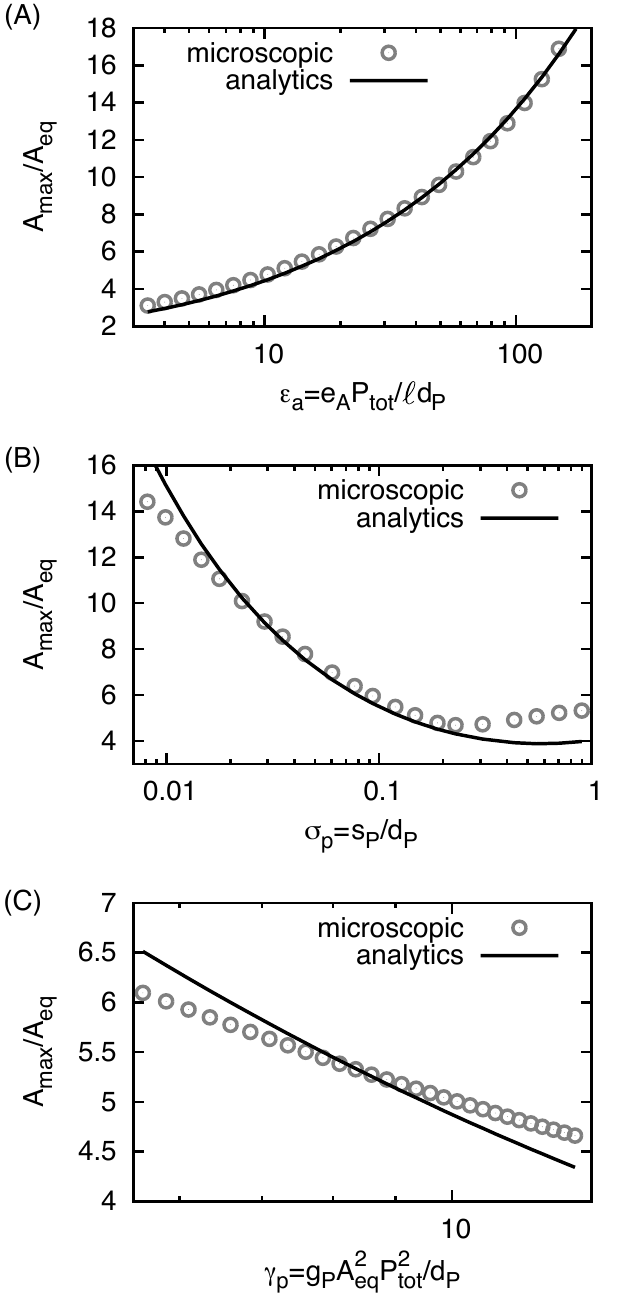}}
\caption{Results for the auxin amplitude $A_{\text{max}}$. The auxin amplitude $A_{\text{max}}$ varies during  a polarization pulse or front with its independent kinetic parameters (A) transport efficiency of auxin $\epsilon_a=e_A P_{\text{tot}}/\ell d_P$, (B) basal attachment rate $\sigma_p=s_P/d_P$ and, (B) enhanced attachment rate $\gamma_p=g_P A_{\text{eq}}^2 P_{\text{tot}}^2/d_P$. Shown are results from the numerical integration of the microscopic equations and the analytic expression multiplied by an overall factor of $2.3$. 
We considered $\sigma_p=s_P/d_P\ll1$ such that only less than a third of all PIN proteins occupy each membrane in the resting state. $\epsilon_p=e_A P_{\text{tot}}/\ell d_P> 1$ as auxin permeability $e_A P_{\text{tot}}=1.4\mu m/s$ \cite{Delbarre:1996p1372,Swarup:2005p429} is roughly larger than endosome cycling by active transport along a cell's cytoskeleton $\ell s_P$ \cite{Howe:2004p4194}.  We suggest $\gamma_p=g_P A_{\text{eq}}^2P_{\text{tot}}^2/d_P>1$ as protein and auxin numbers might be very large. Finally, assuming literature values of $d_A$  \cite{Rapparini:2002p523}, endosome cycling, and taking cell length of tens of $\mu m$, we used $d_A/d_P=0.2$ to compare our results.  
Each graph shows the variation of a single parameter, while the remaining are kept constant at $d_A/d_P=0.2$,  $e_A P_{\text{tot}}/\ell d_P=10$, $g_P A_{\text{eq}}^2 P_{\text{tot}}^2/d_P=12$, and $s_P/d_P=0.2$. }
\label{fig_amax}
\end{figure}
Identifying polarization dynamics as an excitable medium enables us to go beyond numerical integration of the microscopic equations and analytically compute the auxin amplitude of a polarization pulse or front. To this end we employed singular perturbation theory \cite{ORTOLEVA:1975p543,CASTEN:1975p3477} on a polarization pulse. Here we explain the outline of the calculation, a detailed derivation is provided in  Supplemental I. The whole polarization pulse can be subdivided into four regions as shown in fig.~\ref{fig_polar}~(A) and fig.~\ref{fig_null}~(A), first the front and back where the auxin concentration is nearly constant and only the number of PIN proteins changes significantly. Second the excited and the refractory domain during which the efflux facilitator concentration follows approximately the nullcline and hence only changes according to the nullclines' variation with the auxin concentration while the auxin concentration itself varies profoundly.  Therefore, all four regions are governed to good approximation by just a single non-linear equation, the one of the PIN protein dynamics or the one of auxin dynamics, respectively. Unfortunately, even each single continuum equation is not analytically solvable. The model equations (\ref{eqn_acont}, \ref{eqn_prcont}) are therefore linearized around the stable nullclines yielding two sets of equations, one left to the unstable manifold and one right to it. By this linearization we overestimated the PIN dynamics close to the unstable manifold which leads to smaller auxin amplitudes than those resulting from the integration of the microscopic equations (\ref{eqn_a}-\ref{eqn_pl}). An algebraic solution is obtained by imposing a traveling wave Ansatz $A(x-vt)$, $P_r(x-vt)$ and solving all four equations under the condition of differential continuity at their intersections.  The calculation yields a closed expression for the  auxin amplitude which captures the role of the underlying kinetic parameters:
\begin{equation}
\frac{A_{\text{max}}^2}{A_{\text{eq}}^2}=\frac{\rho}{2}\Bigg(1+\sqrt{1+\frac{4^2}{\beta^2}\Bigg[1+\sin\big(\tfrac{\phi}{3}\big)-\tfrac{\cos\big(\tfrac{\phi}{3}\big)}{\sqrt{3}}\Bigg]}\Bigg),
\label{eqn_amax}
\end{equation}
where we abbreviated $\rho=4(1+2\sigma_p)/\gamma_p$, $\beta=16 \frac{1}{\epsilon_a}\frac{\sigma_p}{(1+\sigma_p)}$, and $\phi=\tan^{-1}(-3\sqrt{3}\beta^2,\sqrt{2^{14}-3^3\beta^4}/)$. The analytic result is compared to the numeric integration of the microscopic equations (\ref{eqn_a}-\ref{eqn_pl}) over a broad range of the kinetic parameters in fig.~\ref{fig_amax}. Due to the linearization of the equations the algebraic amplitudes are too low. Fitting the analytic to the numeric results yields an overall factor $2.3$. This constant factor does not depend on the specific parameter range. Considering the amount of approximations that entered the calculation, the analytic result captures very well the dependence on the kinetic parameters over orders of magnitudes.  To obtain insight in how the magnitude of the kinetic parameters determines the amplitude, we simplified eq.~(\ref{eqn_amax}) further. Following our considerations on the size of the kinetic parameters in fig.~\ref{fig_amax}, $\beta$ is smaller than one. Hence, expanding for small $\beta$ gives,
\begin{equation}
A_{\text{max}}^2\propto\frac{2\rho}{\beta}\propto\frac{e_AP_{\text{tot}}}{\ell d_P}\frac{1}{\tfrac{g_PP^2_{\text{tot}}A^2_{\text{eq}}}{d_P}}\frac{(1+\tfrac{s_P}{d_P})(1+2\tfrac{s_P}{d_P})}{\tfrac{s_P}{d_P}}\;.
\label{eqn_shamax}
\end{equation}
As the amplitude arises due to auxin inflow from neighboring cells $A_{\text{max}}$ increases with the rescaled transport efficiency $e_AP_{\text{tot}}/\ell d_P$. Auxin is accumulated in a cell until the cell reaches its almost fully polarized state and can efficiently transport auxin onwards, see fig.~\ref{fig_null}~(B). Hence, accumulation time and auxin amplitude $A_{\text{max}}$ inversely depend on the basal PIN cycling rate $s_P/d_P$ and the enhanced attachment rate $g_P P_{\text{tot}}^2A_{\text{eq}}^2/d_P$.  Synthesis and degradation of auxin do not contribute to the amplitude in this first order approximation as their impact on the transport is very small. Note, that the amplitude like all other pulse and front characteristics is independent of the amount of supplied auxin, a general property of excitable media. Applying the above simplification to the resulting analytic expressions for the velocity of the auxin pulse, see Supplemental eq.~(S7), (S8), yields,
\begin{equation}
v\propto e_A P_{\text{tot}}\;.
\label{eqn_v}
\end{equation} 
This result reveals that the pulse of front velocity is dominated by membrane permeability of auxin facilitated by PIN proteins which is just the product of transport efficiency and number of PIN proteins per cell. The prefactor in eq.~(\ref{eqn_v}) $v/e_AP_{\text{tot}}$ can be estimated from comparison with numeric integration of the microscopic equations as shown in fig.~\ref{fig_amax} to be in the range of  $6\cdot10^{-5}-6\cdot10^{-4}$ depending on the remaining kinetic parameters. 
%%%%%%%%%%%%%%%%%%%%%%%%%%%%%%%%%%%%%%%%%%%%%%%%
\section{Discussion}
%%%%%%%%%%%%%%%%%%%%%%%%%%%%%%%%%%%%%%%%%%%%%%%%
We have shown that some prominent aspects of auxin and PIN dynamics can be inferred from a simple mathematical model. Our model predicts the transition of ground meristem cells from a non-polarized stable state to a polarized stable state of constant auxin flow. This development occurs by a traveling wave front triggered by a continuous inflow of auxin from the outer epidermal layer, in accordance with experimental observations \cite{Scarpella:2006p352}. Each cell is a bistable excitable medium. Excitations from one state to the other can be induced by supply of auxin and crucially depend on the spatial gradients in auxin and polar PIN concentrations between cells. The amplitude of auxin in the wave front and the polarization of the stable states is cast in analytical expressions concordant with numerical integration of our microscopic equations.

The microscopic equations underlying our model for auxin and PIN dynamics are defined such that all relevant biological processes are included while a minimum of assumptions on their actual kinetics entered. To this end only linear synthesis, degradation and transport etc.~is considered as a first order approximation of any kind of kinetics. However, the model is robust against alteration of the linear relationships, as is illustrated in Supplemental II. For example, extending the cell to cell auxin transport  to account for Michaelis-Menten kinetics preserves the form of the nullclines and the dynamics of the wave. In our model only the feedback of auxin flow on the attachment of PIN proteins enters non-linearly.  A linear growth of the enhanced attachment rate with the auxin flow cannot lead to a propagating front, as such a model does not exhibit two stable fixed points. Only auxin flow exponents higher than one show these properties. However, the exact value of the exponent  does not affect the form and dynamics of the traveling wave, again confirming the robustness of our assumptions. 

Recently, models for auxin and PIN dynamics were developed proposing that the auxin concentration in the neighboring cell  feeds back onto an enhanced attachment rate of PIN proteins \cite{Jonsson:2006p421,Smith:2006p585,Merks:2007p3119,Ibanes:2009p4064} in contrast to canalization models, where the net auxin flux governs the feedback. These concentration driven models exhibit a static state of spatially ordered auxin maxima with PIN proteins polarized towards these auxin maxima  \cite{Jonsson:2006p421}. This behavior arises as concentration driven feedback changes the non-linear character of auxin and PIN dynamics. These models generally exhibit only a single stable, resting fixed point \cite{Newell:2008p356,Sahlin:2009p2675}. The polarization due to auxin supply observed in these models \cite{Merks:2007p3119} arises due to an evolved relaxation into the stable, resting state. Hence, the polarization is only temporary and, for instance, the amount of polarization and the velocity of the polarization front depend crucially on the amount of auxin supply. This is in contrast to our minimal canalization model, where all polarization characteristics are only governed by the kinetic parameters. The amount of auxin supply in an excitable medium only regulates if a pulse or front is excited or not. Hence, these qualitative differences may help to distinguish between the different models experimentally. 

The role of all kinetic processes during the dynamic rearrangement of PIN and auxin in cells becomes explicit when examining the very front of the polarization in a microscopic scenario as illustrated in fig.~\ref{fig_null}~(B). The almost fully polarized cell at the peak of the front carries a lot of auxin molecules that are invading the next yet non-polar cell in the direction of polarization with a rate mainly governed by the cell-to-cell transport efficiency $e_A$. To successfully transfer the accumulating auxin onwards the PIN proteins in the yet non-polar cell have to rearrange to facilitate directed transport. However, the endosome cycling  $s_P$, $g_P$ by which the membrane bound PIN proteins reach the cell membrane is very slow. Hence, the attachment rate of the efflux facilitators forms a bottleneck that piles up more and more auxin in a cell, that is slowly increasing the amount of PIN proteins facing the direction of transport. Heuristically, an auxin pulse forms due to a traffic jam caused by the slow cycling of the efflux facilitators, as given by eq.~(\ref{eqn_shamax}). As an equilibrium concentration of PIN proteins is always embedded in each membrane ready to transport auxin, the magnitude of the velocity of a polarization front or pulse is set by the cell-to-cell transport efficiency, see eq.~(\ref{eqn_v}). The other kinetic parameters only slightly modulate the velocity. The PIN attachment rates, $g_P$, $s_P$, and detachment rate, $d_P$, on the other hand determine the number of PIN proteins accumulated at the membrane in the polar stable state, see  eq.~(\ref{eqn_polar}). The polarization grows with the enhanced attachment rate, the strength of the feedback, $g_P$. On the contrary, basal PIN cycling $s_P/d_P$ intensifies the competition between opposing membranes decreasing the amount of PIN proteins in the direction of polarization. 

The result of our analytic expressions for the PIN concentration in the polar state with constant auxin flux in eq.~(\ref{eqn_polar}) and the auxin amplitude at the very head of the polarization front eq.~(\ref{eqn_amax}) enable estimates of the underlying kinetic rates by identifying and measuring these observables in future experiments. Existing experimental results by Scarpella et al.~\cite{Scarpella:2006p352} permit an estimate of the polarization front velocity in the range $v=10^{-4}-10^{-3}\mu m/s$, in accordance with our estimate for the velocity $v=8\cdot10^{-5}-8\cdot10^{-4}\mu m/s$, resulting from the fitted pre-factor in eq.~(\ref{eqn_v}) and the literature value of auxin permeability $e_AP_{\text{tot}}$ \cite{Delbarre:1996p1372,Swarup:2005p429}. An quantitative estimate of PIN polarity from the same existing data is to best of our knowledge yet unfeasible as a reference for the protein number is absent. This could be overcome by new experiments, which could also aim at the auxin kinetic rates. Unlike PIN which is readily GFP tagged, auxin is not directly detectable and quantification of its amount can only occur via indirect methods. Recently, measurements of deuterated auxin improved \cite{Ljung:2005p397} making experiments with exogenously applied auxin conceivable. In such setups one should, however, keep in mind that exogenous auxin mixes with endogenous, non-labelled auxin, decreasing the observed amplitude. The position of the auxin peak can easily be located as it should be accumulated in front of those cells with the largest amount of PIN proteins at the corresponding membrane. Measurements of the amount of PIN proteins in polarized cells via GFP tagging could not only disclose the basal endosome cycling rate but also the magnitude of the feedback between auxin flow and PIN dynamics.    

The occurrence of bipolar cells has stimulated previous theoretical models introducing a hypothetical new molecule  \cite{Feugier:2006p477} or moving auxin sources \cite{RollandLagan:2005p476}. Our model, however, readily predicts the occurrence of bipolar cells along a one-dimensional strand of cells at the site of continuous auxin inflow. These cells show a high concentration of PIN proteins on either membrane, a balanced outcome of the competition for PIN between both membranes. This state is not a statically stable but dynamically driven by the supply of auxin. Transferring this observation to the two-dimensional layer of ground meristem cells during vein initiation may explain the bipolar cells observed experimentally \cite{Scarpella:2006p352,Wenzel:2007p1028,Bayer:2009p4590,Koenig:2009p4588}. In two dimensions several membranes can compete, yielding also triple polar cells or theoretically higher orders of polarity. However, in biological cells not all cell membranes may have the same number of PIN proteins to enter the competition. Those with fewer initial PIN proteins will become the site of PIN drain, decreasing the number of successful polarization fronts. Experiments indicate that mechanical cues might favor certain membranes \cite{Hamant:2008p627} paving the way of a polarization front and hence the position of veins and vein loops.

In summary we analyzed a canalization model to explain how auxin and PIN dynamics polarize the distribution of PIN proteins during early vein initiation in the ground meristem. Each cell is found to be bistable considering their PIN distribution in the membrane. Polarization occurs in a traveling front as auxin is supplied to a single cell mimicking the auxin inflow from the outer epidermal cell layer. The driver in this transition is the spatial gradient of auxin and polar PIN concentration between cells. The key idea in this polarization is the positive feedback between hormone auxin and its own efflux facilitator PIN. An idea that might be inspiring in other developmental processes in animals where tissue is polarized as, for example, in planar cell polarity \cite{Lawrence:2007p4177}. As the polarization front can travel in opposite directions from its initiation site, a bipolar cell, the up to now puzzling occurrence of closed vein loops can be resolved. Furthermore, our nonlinear analysis enables the calculation of exact analytical expression for the polarization front. Therefore, our new quantitative predictions for the PIN polarization driven by auxin flow establishes a basis to determine the kinetic parameters underlying the transport of auxin and may therewith have far-reaching impacts on the understanding of the developmental processes and their differences in plant species, to perceive the fundamental patterns of leaf veins or phyllotaxis and learn how environmental conditions alter these.
%%%%%%%%%%%%%%%%%%%%%%%%%%%%%%%%%%%%%%%%%%%%%%%%
\section*{Acknowledgements}
%%%%%%%%%%%%%%%%%%%%%%%%%%%%%%%%%%%%%%%%%%%%%%%%
The authors thank Eric M. Kramer for fruitful discussions. Financial support of the German Excellence Initiative via the program "Nanosystems Initiative Munich (NIM)" and of the LMUinnovativ project "Functional Nanosystems (FuNS)" is gratefully acknowledged. K.A.~acknowledges funding by the Studienstiftung des deutschen Volkes. 
%%%%%%%%%%%%%%%%%%%%%%%%%%%%%%%%%%%%%%%%%%%%%%%%

%%%%%%%%%%%%%%%%%%%%%%%%%%%%%%%%%%%%%%%%%%%%%%%%
% Supplemental
%%%%%%%%%%%%%%%%%%%%%%%%%%%%%%%%%%%%%%%%%%%%%%%%
\cleardoublepage


\section*{Supplementary material}

# Supplementary Information: Quantitative Predictions on Auxin-Induced Polar Distribution of PIN Proteins during Vein Formation in Leaves

Karen Alim and Erwin Frey

*Arnold Sommerfeld Center for Theoretical Physics and Center for NanoScience,*

*Ludwig-Maximilians-Universität, Theresienstr. 37, D-80333 München, Germany*

## I. AUXIN AMPLITUDE COMPUTATION

Starting from the microscopic definition of a model for the polarization of PIN distributions by auxin flow, see Eq. (1), (2) and (3), an exact analytical expression for the auxin amplitude and the velocity of the polarization pulse is derived. To this end a continuum limit is performed and singular perturbation theory is employed as described in the following.

Initially, polarization is defined by three equations governing the dynamics of auxin concentration $A$ and the amount of PIN efflux facilitators on the membrane on the right $P_r$ or on the left $P_l$ hand side of the cell. Assuming the symmetry of the system is broken such that polarization evolves to the right, the concentration of efflux facilitators on the left hand side membrane of every cell does not change significantly with time if $s_P/(d_P + 2s_P) \ll 1$, we therefore assume $dP_l(n)/dt = 0$. As flow proceeds to the right $J(n) > 0 \; \forall n$, the amount of PIN proteins on the left hand side membrane amounts in its stationary state to $P_l(n) = s_P(1 - P_r(n))/(d_P + s_P)$. Substituting this result in the remaining dynamic equations the system is described by two components only, $A(n)$ and $P_r(n)$. We derive continuum equations by setting $n \to x$, $n + 1 \to x + \ell$, and $n - 1 \to x - \ell$. When the wavelength as the length scale on which the pulse evolves is considerably larger than the cell length $\ell$, a time scale separation occurs which makes higher order terms negligible small. As observed in the simulations shown in Fig. 2 the wavelength of a single pulse is of the order of tens of cells justifying a Taylor expansion in $x$. To describe the characteristics observed in the microscopic equations with continuous equations only zeroth order terms and a single second order term in the PIN dynamics are required,

$$\frac{\partial}{\partial t}\frac{A(x,t)}{A_{\text{eq}}} = d_A\left(1 - \frac{A(x,t)}{A_{\text{eq}}}\right) - \frac{e_A P_{\text{tot}}}{\ell}\frac{1 + 2\frac{s_P}{d_P}}{1 + \frac{s_P}{d_P}}\ell\frac{\partial}{\partial x}\left[\left(\frac{P_r(x,t)}{P_{\text{tot}}} - \frac{\frac{s_P}{d_P}}{1 + 2\frac{s_P}{d_P}}\right)\frac{A(x,t)}{A_{\text{eq}}}\right], \quad \text{(S1)}$$

$$\frac{\partial}{\partial t}\frac{P_r(x,t)}{P_{\text{tot}}} = -\frac{g_P P_{\text{tot}}^2 (1+2\frac{s_P}{d_P})^2 A^2(x,t)}{(1+\frac{s_P}{d_P})^3}\left(\frac{P_r(x,t)}{P_{\text{tot}}}-\frac{\frac{s_P}{d_P}}{1+2\frac{s_P}{d_P}}\right)$$

$$\times\left(\frac{P_r(x,t)}{P_{\text{tot}}}-\frac{1+3\frac{s_P}{d_P}-(1+\frac{s_P}{d_P})\sqrt{1-\frac{4(d_P+2s_P)}{g_P P_{\text{tot}}^2 A^2(x)}}}{2(1+2\frac{s_P}{d_P})}\right)$$

$$\times\left(\frac{P_r(x,t)}{P_{\text{tot}}}-\frac{1+3\frac{s_P}{d_P}+(1+\frac{s_P}{d_P})\sqrt{1-\frac{4(d_P+2s_P)}{g_P P_{\text{tot}}^2 A^2(x,t)}}}{2(1+2\frac{s_P}{d_P})}\right)$$

$$-\frac{g_P P_{\text{tot}}^2 \frac{s_P^2}{d_P^2}}{(1+\frac{s_P}{d_P})^3}\left(\frac{P_r(x,t)}{P_{\text{tot}}}-1\right)\left\{\ell\frac{\partial}{\partial x}\left[\left(\frac{P_r(x,t)}{P_{\text{tot}}}-1\right)A(x,t)\right]\right\}^2 . \tag{S2}$$

In these expressions the degree of nonlinearity is still too high to obtain analytical results by use of singular perturbation theory. Therefore, the continuum equations are further simplified by approximating them right and left of the nullcline embedding the unstable fixed point, $P_r(x) = \frac{d_P+3s_P}{2(d_P+2s_P)} - \frac{d_P+s_P}{2(d_P+2s_P)}\sqrt{1-\frac{4(d_P+2s_P)}{g_P P_{\text{tot}} A^2(x)}}$, which to good approximation resembles the separatrix between the areas of attraction of the stable fixed points.

Left of the nullcline the reaction terms of the continuum equations, those terms without spatial or temporal derivatives, are expanded around the stable nullcline embedding the stable resting state ($A/A_{\text{eq}} = 1, P_r/P_{\text{tot}} = s_P/(d_P + 2s_P)$). The spatial derivative terms are simplified by discarding the spatial derivative of auxin whose factor $P_r(x)/P_{\text{tot}} - s_P/(d_P + 2s_P)$ turns the whole term negligible small close to the stable nullcline, resulting in,

$$\frac{\partial}{\partial t}\frac{A(x,t)}{A_{\text{eq}}} = d_A\left(1-\frac{A(x,t)}{A_{\text{eq}}}\right) - \frac{e_A P_{\text{tot}}}{\ell}\frac{1+2\frac{s_P}{d_P}}{1+\frac{s_P}{d_P}}\frac{A(x,t)}{A_{\text{eq}}}\ell\frac{\partial}{\partial x}\frac{P_r(x,t)}{P_{\text{tot}}}, \tag{S3}$$

$$\frac{\partial}{\partial t}\frac{P_r(x,t)}{P_{\text{tot}}} = -\frac{(d_P+2s_P)}{(1+\frac{s_P}{d_P})}\left(\frac{P_r(x,t)}{P_{\text{tot}}}-\frac{\frac{s_P}{d_P}}{1+2\frac{s_P}{d_P}}\right)$$

$$-\frac{g_P P_{\text{tot}}^2 \frac{s_P^2}{d_P^2}}{(1+\frac{s_P}{d_P})^3}\left(\frac{P_r(x,t)}{P_{\text{tot}}}-1\right)\left\{\ell\frac{\partial}{\partial x}\left[\left(\frac{P_r(x,t)}{P_{\text{tot}}}-1\right)A(x,t)\right]\right\}^2 . \tag{S4}$$

Right of the nullcline embedding the unstable fixed point the reaction term of the efflux facilitator dynamics is expanded around the polar fixed point ($A/A_{\text{eq}} = 1, P_r/P_{\text{tot}} = \frac{d_P+3s_P}{2(d_P+2s_P)} + \frac{d_P+s_P}{2(d_P+2s_P)}\sqrt{1-\frac{4(d_P+2s_P)}{g_P A_{\text{eq}}^2 P_{\text{tot}}^2}}$). Here, the spatial derivative terms contributing to the PIN protein dy-

namics become negligible small and are therefore discarded, yielding,

$$
\frac{\partial}{\partial t}\frac{A(x,t)}{A_{\text{eq}}} = d_A\left(1-\frac{A(x,t)}{A_{\text{eq}}}\right) - \frac{e_A P_{\text{tot}}}{\ell}\frac{1+2\frac{s_P}{d_P}}{1+\frac{s_P}{d_P}}\ell\frac{\partial}{\partial x}\left[\left(\frac{P_r(x,t)}{P_{\text{tot}}}-\frac{\frac{s_P}{d_P}}{1+2\frac{s_P}{d_P}}\right)\frac{A(x,t)}{A_{\text{eq}}}\right]\tag{S5}
$$

$$
\frac{\partial}{\partial t}\frac{P_r(x,t)}{P_{\text{tot}}} = -\frac{g_P P_{\text{tot}}^2 A^2(x,t)}{2(1+\frac{s_P}{d_P})}\left(1-\frac{4(d_P+2s_P)}{g_P P_{\text{tot}}^2 A^2(x,t)}+\sqrt{1-\frac{4(d_P+2s_P)}{g_P P_{\text{tot}}^2 A^2(x,t)}}\right)
$$
$$
\times\left(\frac{P_r(x,t)}{P_{\text{tot}}}-\frac{1+3\frac{s_P}{d_P}+(1+\frac{s_P}{d_P})\sqrt{1-\frac{4(d_P+2s_P)}{g_P P_{\text{tot}}^2 A^2(x,t)}}}{2(1+2\frac{s_P}{d_P})}\right).\tag{S6}
$$

Based on these sets of equations the amplitude and the velocity of a polarization pulse are calculated employing singular perturbation theory [1, 2]. The singular perturbation approach becomes applicable to a two component system performing a pulse if the pulse can be separated in different regions which comply either of the following restrictions. Either to good approximation one component is constant while the other changes rapidly. Or if both components change simultaneously one of the components should follow a nullcline. Then the two coupled nonlinear equations decouple in each region and only one differential equation remains to be solved. Assuming a traveling wave ansatz $z = x - vt$, where $v$ defines the velocity of the wave, the partial differential equations simplify to ordinary differential equations. These differential equations remain to be solved under the condition of continuity and differential continuity at the nullcline embedding the unstable fixed point which separates the two cases Eqs. (S3, S4) and Eqs. (S5, S6).

The trajectory of an auxin pulse can be subdivided into four regions, first a wave front and back, where the auxin concentration is approximately constant $A_{\text{max, min}}$, while the PIN protein concentration changes rapidly and second an excited and a refractory region during which the PIN concentration follows the stable nullclines. Several boundary conditions arise from the requirement of continuity and differential continuity. Considering the efflux facilitator dynamics continuity requires that PIN concentrations during pulse front and back governed by Eqs. (S4) and (S6) merge into the nullclines defining refractory and excited domain $P_{r,\text{front}}(z \to \infty) = P_{r,\text{refrac}}$, $P_{r,\text{front}}(z \to -\infty) = P_{r,\text{excite}}$ and $P_{r,\text{back}}(z \to -\infty) = P_{r,\text{refrac}}$, $P_{r,\text{back}}(z \to \infty) = P_{r,\text{excite}}$. As the separating nullcline is crossed during wave front and back additionally continuity and differential continuity is compulsory for efflux facilitators at the position of the nullcline embedding the unstable fixed point $z_{\text{sep}}$. The auxin concentration evolving during excited Eq. (S5) and refractory region Eq. (S3) has to reach the constant auxin concentration of pulse front and back $A_{\text{max}}$, $A_{\text{min}}$ at distinct points

in evolution $z = z_{\text{front}}$ and $z_{\text{back}}$, resulting in the boundary conditions $A_{\text{refrac}}(z_{\text{front}}) = A_{\text{max}} = A_{\text{excite}}(z_{\text{front}})$, $A_{\text{refrac}}(z_{\text{back}}) = A_{\text{min}} = A_{\text{excite}}(z_{\text{back}})$, $dA_{\text{refrac}}(z_{\text{front}})/dz = dA_{\text{excite}}(z_{\text{front}})/dz$, and $dA_{\text{refrac}}(z_{\text{back}})/dz = dA_{\text{excite}}(z_{\text{back}})/dz$. Two of those boundary conditions yield equations that solve for the auxin amplitude $A_{\text{max}}$ and the pulse velocity $v$. First the condition of differential continuity of the PIN concentration at the separatrix $\partial P_r(z_{\text{sep}})/\partial z$ results in,

$$v = \frac{4\ell d_P \frac{s_P}{d_P}}{1 + \frac{s_P}{d_P}} \sqrt{\frac{\frac{A_{\text{max}}^8}{\rho^4}\left(1 + \frac{\rho}{A_{\text{max}}^2}\right)^2 \left(1 + \sqrt{1 - \frac{\rho}{A_{\text{max}}^2}}\right)^4}{1 + \frac{4A_{\text{max}}^4}{\rho^2}\left(1 + \frac{\rho}{A_{\text{max}}^2}\right)\left(1 + \sqrt{1 - \frac{\rho}{A_{\text{max}}^2}}\right)^2}} \,, \tag{S7}$$

where we abbreviated $\rho = 4(d_P + 2s_P)/g_P P_{\text{tot}}^2$. Second differential continuity of auxin left and right of a wave front or back $A_{\text{refrac}}(z_{\text{front}}) = A_{\text{max}} = A_{\text{excite}}(z_{\text{front}})$ yields,

$$v = \frac{e_A P_{\text{tot}}}{4}\left(1 + \sqrt{1 - \frac{\rho}{A_{\text{max}}^2}}\right). \tag{S8}$$

These two equations together result in a quartic equation for the amplitude of the auxin pulse, evaluated to the expression given in eq. (8). Reentering this result into one of the defining equations above yields an analytic result for the velocity of an auxin pulse.

## II. COMPARISON TO ALTERNATIVE MICROSCOPIC TRANSPORT MODELS

When defining our microscopic equations Eqs. (1 - 3) we included all contributing processes but considered a minimum of assumptions on the kinetics. In this spirit all processes are modeled by linear relations as the first order term of any kind of underlying kinetics. The only exception is the enhanced attachment of PIN proteins, which is the point where non-linearity enters the microscopic model triggering the non-linear effect of a traveling pulse or front. Including further non-linearities renders the microscopic models intractable for analytical calculations. However, the analysis of our minimal model has revealed the key characteristics for polarization to be two stable fixed points accompanied by an unstable manifold. Excitations beyond this unstable manifold then lead to the development of a wave pulse or the relaxation to the polar stable fixed point. With this knowledge we can assess more evolved microscopic models by comparing their non-linear characteristics such as the nullclines to the minimal model. As is shown in the following,

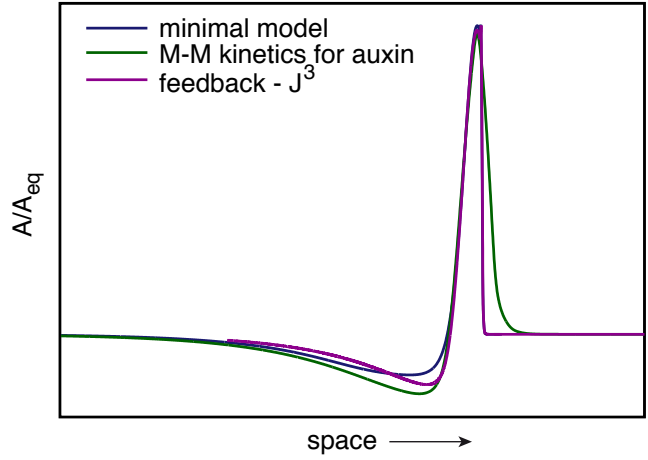

FIG. S1: Comparing the polarization pulse for different extended models to our minimal model yields no difference in their characteristics confirming the robustness of our linear approximations. Parameters values as follows. Minimal model $d_A/d_P = 0.2$, $s_P/d_P = 0.2$, $e_A P_{\text{tot}}/\ell d_P = 10$, and $g_P A_{\text{eq}}^2 P_{\text{tot}}^2/d_P = 12$. Michaelis-Menten kinetics for active auxin transport $d_A/d_P = 0.2$, $s_P/d_P = 0.2$, $e_A P_{\text{tot}}/\ell d_P A_{\text{eq}} = 8$, $g_P P_{\text{tot}}^2/d_P = 70$, and $k_A/A_{\text{eq}} = 1.6$. Feedback to the power of three $d_A/d_P = 0.2$, $s_P/d_P = 0.2$, $e_A P_{\text{tot}}/\ell d_P A_{\text{eq}} = 10$, and $g_P P_{\text{tot}}^3 A_{\text{eq}}^3/d_P = 18$.

exemplarily changing linear terms of the minimal model into non-linear terms results only in slight changes of the characteristics confirming the robustness of a first order assumption in the minimal model. If not explicitly stated otherwise we assumed $dP_l(n)/dt = 0$ to calculate the fixed points.

### A. Michaelis-Menten kinetics for active auxin transport

One may assume that a Michaelis-Menten mechanism describes the active transport of a substrate auxin by an enzyme presented by the PIN proteins [3, 4]. Then, in the definition of the net auxin flow the number of transported auxin molecules is represented by a Hill function with Michaelis-Menten constant $k_A$. For reasons of completeness we state the full set of microscopic equations:

$$\frac{d}{dt}A(n) = s_A - d_A A(n) - \frac{e_A}{\ell}\left[J(n) - J(n-1)\right], \tag{S9}$$

$$\frac{d}{dt}P_r(n) = -d_P P_r(n) + s_P P_b(n) + g_P J^2(n)\theta(J(n))P_b(n), \tag{S10}$$

$$\frac{d}{dt}P_l(n) = -d_P P_l(n) + s_P P_b(n) + g_P J^2(n-1)\theta(-J(n-1))P_b(n), \tag{S11}$$

$$J(n) = \frac{A(n)}{A(n) + k_A}P_r(n) - \frac{A(n+1)}{A(n+1) + k_A}P_l(n+1). \tag{S12}$$

These equations yield in accordance with the minimal model three fixed points, one resting state at $(A/A_{\text{eq}} = 1, P_r/P_{\text{tot}} = s_P/(d_P + 2s_P))$ and a pair of stable and unstable constant current fixed points at $(A/A_{\text{eq}} = 1, P_r/P_{\text{tot}} = \frac{d_P + 3s_P}{2(d_P + 2s_P)} \mp \frac{d_P + s_P}{2(d_P + 2s_P)} \sqrt{1 - (1 + k_A)^2 \frac{4(d_P + 2s_P)}{g_P P_{\text{tot}}^2}})$. The model displays very similar nullclines and the same dynamics as the minimal model as exemplified in Fig. S1. If the new parameter $k_A$ lies outside its range $k_A \leq -1 - \sqrt{\frac{g_P P_{\text{tot}}^2}{4(d_P + 2s_P)}}$ and $k_A \geq -1 + \sqrt{\frac{g_P P_{\text{tot}}^2}{4(d_P + 2s_P)}}$ only a single stable fixed point occurs and no polarization can be observed.

## B. Feedback - power of the current

In the minimal model we take the feedback of auxin flow on the enhanced attachment of PIN proteins to enter with a power of two. In general one could assume any kind of power,

$$\frac{d}{dt}A(n) = s_A - d_A A(n) - \frac{e_A}{\ell}\left[J(n) - J(n-1)\right], \tag{S13}$$

$$\frac{d}{dt}P_r(n) = -d_P P_r(n) + s_P P_b(n) + g_P J^k(n)\theta(J(n))P_b(n), \tag{S14}$$

$$\frac{d}{dt}P_l(n) = -d_P P_l(n) + s_P P_b(n) + g_P J^k(n-1)\theta(-J(n-1))P_b(n), \tag{S15}$$

$$J(n) = A(n)P_r(n) - A(n+1)P_l(n+1). \tag{S16}$$

Considering only integer powers for simplicity, we find the following. For $k = 1$ the above equations display the resting fixed point at $(A/A_{\text{eq}} = 1, P_r/P_{\text{tot}} = P_l/P_{\text{tot}} = s_P/(d_P + 2s_P))$ and a polar fixed point at $(A/A_{\text{eq}} = 1, P_l/P_{\text{tot}} = s_P/g_P A_{\text{eq}} P_{\text{tot}}, P_r/P_{\text{tot}} = 1 - (s_P + d_P)/g_P A_{\text{eq}} P_{\text{tot}})$. The resting state is the only fixed point and stable for $g_P A_{\text{eq}} P_{\text{tot}} < d_P + 2s_P$, otherwise the polar fixed point is stable and the resting state turns unstable. No excited polarization can occur in either case. However, if $k > 1$ the equations display a set of stable and unstable fixed points in addition to the resting stable fixed point. For any $k > 1$ one recovers the dynamics observed for $k = 2$ in the minimal model as exemplified in Fig. S1 for the cases $k = 3$. This observation is in accordance with results of Ref. [5], which stated vein patterns for any feedback function obeying to first order a higher power than $k = 1$. For $k = 2$ we observed that the pair of a stable and an unstable polar fixed point occurs only for $g_P A_{\text{eq}}^2 P_{\text{tot}}^2/d_P \geq 4(1 + 2s_P/d_P)$, similar rules apply for $k = 3$ or higher.


%%%%%%%%%%%%%%%%%%%%%%%%%%%%%%%%%%%%%%%%%%%%%%%%

\begin{thebibliography}{56}

\bibitem{Petrasek:2009p4589}
J.~Petrasek, J.~Friml, Development \textbf{136}(16), 2675 (2009)

\bibitem{Friml:2002p1835}
J.~Friml, J.~Wi\'sniewska, E.~Benkov\'a, K.~Mendgen, K.~Palme, Nature
  \textbf{415}(6873), 806 (2002)

\bibitem{Marchant:1999p4592}
A.~Marchant, J.~Kargul, S.T. May, P.~Muller, A.~Delbarre, C.~Perrot-Rechenmann,
  M.J. Bennett, EMBO J \textbf{18}(8), 2066 (1999)

\bibitem{Reinhardt:2003p348}
D.~Reinhardt, E.R. Pesce, P.~Stieger, T.~Mandel, K.~Baltensperger, M.~Bennett,
  J.~Traas, J.~Friml, C.~Kuhlemeier, Nature \textbf{426}(6964), 255 (2003)

\bibitem{Sieburth:1999p3331}
L.E. Sieburth, Plant Physiol. \textbf{121}(4), 1179 (1999)

\bibitem{Went:1933p3768}
F.A.F.C. Went, Naturwissenschaften \textbf{21}, 1 (1933)

\bibitem{RUBERY:1973p3312}
P.H. Rubery, A.R. Sheldrake, Nature - New Biol. \textbf{244}(139), 285 (1973)

\bibitem{Goldsmith:1981p503}
M.H.M. Goldsmith, T.H. Goldsmith, M.H. Martin, Proc. Natl. Acad. Sci. USA
  \textbf{78}(2), 976 (1981)

\bibitem{MITCHISON:1980p3359}
G.J. Mitchison, Proc. Roy. Soc. Lond. B Bio. \textbf{209}(1177), 489 (1980)

\bibitem{Galweiler:1998p1908}
L.~G\"alweiler, C.H. Guan, A.~M\"uller, E.~Wisman, K.~Mendgen, A.~Yephremov,
  K.~Palme, Science \textbf{282}(5397), 2226 (1998)

\bibitem{Okada:1991p3335}
K.~Okada, J.~Ueda, M.K. Komaki, C.J. Bell, Y.~Shimura, Plant Cell
  \textbf{3}(7), 677 (1991)

\bibitem{Benkova:2003p3360}
E.~Benkov{\'a}, M.~Michniewicz, M.~Sauer, T.~Teichmann, D.~Seifertov{\'a},
  G.~J{\"u}rgens, J.~Friml, Cell \textbf{115}(5), 591 (2003)

\bibitem{Friml:2003p3325}
J.~Friml, A.~Vieten, M.~Sauer, D.~Weijers, H.~Schwarz, T.~Hamann, R.~Offringa,
  G.~J{\"u}rgens, Nature \textbf{426}(6963), 147 (2003)

\bibitem{Petrasek:2006p531}
J.~Petr\'a{\v s}ek, J.~Mravec, R.~Bouchard, J.J. Blakeslee, M.~Abas,
  D.~Seifertov\'a, J.~Wi\'sniewska, Z.~Tadele, M.~Kube{\v s}, M.~{\v
  C}ovanov\'a et~al., Science \textbf{312}(5775), 914 (2006)

\bibitem{SACHS:1969p1017}
T.~Sachs, Ann. Bot. - London \textbf{33}(130), 263 (1969)

\bibitem{MITCHISON:1980p492}
G.J. Mitchison, Proc. Roy. Soc. Lond. B Bio. \textbf{207}(1166), 79 (1980)

\bibitem{MITCHISON:1981p365}
G.J. Mitchison, Philos. Trans. Roy. Soc. B \textbf{295}(1078), 461 (1981)

\bibitem{Sauer:2006p1907}
M.~Sauer, J.~Balla, C.~Luschnig, J.~Wisniewska, V.~Reinohl, J.~Friml,
  E.~Benkova, Genes Dev \textbf{20}(20), 2902 (2006)

\bibitem{Scarpella:2006p352}
E.~Scarpella, D.~Marcos, J.~Friml, T.~Berleth, Genes Dev \textbf{20}(8), 1015
  (2006)

\bibitem{Paciorek:2005p1843}
T.~Paciorek, E.~Za{\v z}imalov\'a, N.~Ruthardt, J.~Petr\'a{\v s}ek, Y.D.
  Stierhof, J.~Kleine-Vehn, D.A. Morris, N.~Emans, G.~J\"urgens, N.~Geldner
  et~al., Nature \textbf{435}(7046), 1251 (2005)

\bibitem{Dhonukshe:2008p367}
P.~Dhonukshe, I.~Grigoriev, R.~Fischer, M.~Tominaga, D.G. Robinson, J.~Hasek,
  T.~Paciorek, J.~Petr{\'a}sek, D.~Seifertov{\'a}, R.~Tejos et~al., P Natl Acad
  Sci USA \textbf{105}(11), 4489 (2008)

\bibitem{Benjamins:2008p381}
R.~Benjamins, B.~Scheres, Annual Rev Plant Biol \textbf{59}, 443 (2008)

\bibitem{Jonsson:2006p421}
H.~J\"onsson, M.G. Heisler, B.E. Shapiro, E.M. Meyerowitz, E.~Mjolsness, Proc.
  Natl. Acad. Sci. USA \textbf{103}(5), 1633 (2006)

\bibitem{Feugier:2005p548}
F.G. Feugier, A.~Mochizuki, Y.~Iwasa, J. Theor. Biol. \textbf{236}(4), 366
  (2005)

\bibitem{Berleth:2007p473}
T.~Berleth, E.~Scarpella, P.~Prusinkiewicz, Trends Plant Sci \textbf{12}(4),
  151 (2007)

\bibitem{Smith:2009p5137}
R.S. Smith, E.M. Bayer, Plant Cell Environ \textbf{32}(9), 1258 (2009)

\bibitem{Wenzel:2007p1028}
C.L. Wenzel, M.~Schuetz, Q.~Yu, J.~Mattsson, Plant J \textbf{49}(3), 387 (2007)

\bibitem{Bayer:2009p4590}
E.M. Bayer, R.S. Smith, T.~Mandel, N.~Nakayama, M.~Sauer, P.~Prusinkiewicz,
  C.~Kuhlemeier, Genes Dev \textbf{23}(3), 373 (2009)

\bibitem{Koenig:2009p4588}
D.~Koenig, E.~Bayer, J.~Kang, C.~Kuhlemeier, N.~Sinha, Development
  \textbf{136}(17), 2997 (2009)

\bibitem{Sawchuk:2007p4455}
M.G. Sawchuk, P.~Head, T.J. Donner, E.~Scarpella, New Phytol \textbf{176}(3),
  560 (2007)

\bibitem{RAVEN:1975p3293}
J.A. Raven, New Phytol. \textbf{74}(2), 163 (1975)

\bibitem{Swarup:2005p429}
R.~Swarup, E.M. Kramer, P.~Perry, K.~Knox, H.M.O. Leyser, J.~Haseloff, G.T.S.
  Beemster, R.~Bhalerao, M.J. Bennett, Nat. Cell. Biol. \textbf{7}(11), 1057
  (2005)

\bibitem{USENKO:1987p3447}
A.S. Usenko, A.G. Zagorodny, Mol. Phys. \textbf{61}(5), 1213 (1987)

\bibitem{byKPaech:1955p3232}
P.~Larsen, \emph{Growth substances in higher plants}, Vol.~3 of \emph{Modern
  methods of plant physiology} (Springer, Berlin, 1955)

\bibitem{WANGERMANN:1981p3137}
E.~Wangermann, G.J. Mitchison, Plant Cell Environ. \textbf{4}(2), 141 (1981)

\bibitem{Kramer:2007p401}
E.M. Kramer, J Exp. Bot. \textbf{59}(1), 45 (2007)

\bibitem{Pollmann:2002p3769}
S.~Pollmann, A.~M{\"u}ller, M.~Piotrowski, E.W. Weiler, Planta \textbf{216}(1),
  155 (2002)

\bibitem{RollandLagan:2005p476}
A.G. Rolland-Lagan, P.~Prusinkiewicz, Plant J \textbf{44}(5), 854 (2005)

\bibitem{Stoma:2008p4203}
S.~Stoma, M.~Lucas, J.~Chopard, M.~Schaedel, J.~Traas, C.~Godin, PLoS Comput
  Biol \textbf{4}(10), e1000207 (2008)

\bibitem{Kramer:2009p4196}
E.M. Kramer, Trends Plant Sci \textbf{14}(5), 242 (2009)

\bibitem{Smith:2006p585}
R.S. Smith, S.~Guyomarc'h, T.~Mandel, D.~Reinhardt, C.~Kuhlemeier,
  P.~Prusinkiewicz, Proc. Natl. Acad. Sci. USA \textbf{103}(5), 1301 (2006)

\bibitem{Merks:2007p3119}
R.M.H. Merks, Y.V. de~Peer, D.~Inz{\'e}, G.T.S. Beemster, Trends Plant Sci
  \textbf{12}(9), 384 (2007)

\bibitem{Ibanes:2009p4064}
M.~Iba{\~n}es, N.~F{\`a}bregas, J.~Chory, A.~Ca{\~n}o-Delgado, P. Natl. Acad.
  Sci. USA \textbf{106}(32), 13630 (2009)

\bibitem{Rapparini:2002p523}
F.~Rapparini, Y.Y. Tam, J.D. Cohen, J.P. Slovin, Plant Physiol.
  \textbf{128}(4), 1410 (2002)

\bibitem{Delbarre:1996p1372}
A.~Delbarre, P.~Muller, V.~Imhoff, J.~Guern, Planta \textbf{198}(4), 532 (1996)

\bibitem{SergeevichZykov:1987p1055}
V.S. Zykov, A.T. Winfree, \emph{Simulation of Wave Processes in Excitable
  Media} (Manchester Univ. Press, Manchester, 1987)

\bibitem{SMikhailov:1991p1044}
A.S. Mikhailov, A.Y. Loskutov, \emph{Foundations of Synergetics I: Distributed
  Active Systems} (Springer Verlag, New York, 1991)

\bibitem{Howe:2004p4194}
C.L. Howe, W.C. Mobley, J Neurobiol \textbf{58}(2), 207 (2004)

\bibitem{ORTOLEVA:1975p543}
P.~Ortoleva, J.~Ross, J. Chem. Phys. \textbf{63}(8), 3398 (1975)

\bibitem{CASTEN:1975p3477}
R.G. Casten, H.~Cohen, P.A. Lagerstrom, Q. Appl. Math. \textbf{32}(4), 365
  (1975)

\bibitem{Newell:2008p356}
A.~Newell, P.~Shipman, Z.~Sun, J Theor Biol \textbf{251}(3), 421 (2008)

\bibitem{Sahlin:2009p2675}
P.~Sahlin, B.~S{\"o}derberg, H.~J{\"o}nsson, J Theor Biol \textbf{258}(1), 60
  (2009)

\bibitem{Ljung:2005p397}
K.~Ljung, A.K. Hull, J.~Celenza, M.~Yamada, M.~Estelle, J.~Normanly,
  G.~Sandberg, Plant Cell \textbf{17}(4), 1090 (2005)

\bibitem{Feugier:2006p477}
F.~Feugier, Y.~Iwasa, J Theor Biol \textbf{243}(2), 235 (2006)

\bibitem{Hamant:2008p627}
O.~Hamant, M.G. Heisler, H.~Jonsson, P.~Krupinski, M.~Uyttewaal, P.~Bokov,
  F.~Corson, P.~Sahlin, A.~Boudaoud, E.M. Meyerowitz et~al., Science
  \textbf{322}(5908), 1650 (2008)

\bibitem{Lawrence:2007p4177}
P.A. Lawrence, G.~Struhl, J.~Casal, Nat Rev Genet \textbf{8}(7), 555 (2007)

\end{thebibliography}
\end{document}